\documentclass[aps,prstab,preprint,groupedaddress]{revtex4}
\usepackage{graphics,epsfig}
\begin{document}

\draft
\title{Epidemics and Dimensionality in Hierarchical Networks}

\author{Da-Fang Zheng$^{1,2,}$
\footnote{Corresponding author.\\ E-mail address:
dfzheng@zjuem.zju.edu.cn.}, P.M. Hui$^{3}$, Steffen Trimper$^{2}$,
and Bo Zheng$^{1,2}$}

\address{$^{1}$ Zhejiang Institute of Modern Physics, Zhejiang University,
Hangzhou 310027, People's Republic of China}
\address{$^{2}$ Fachbereich Physik, Martin-Luther-Universit\"{a}t, D-06099 Halle, Germany}
\address{$^{3}$ Department of Physics, The Chinese University of Hong Kong,
Shatin, Hong Kong, China}

\begin{abstract}
Epidemiological processes are studied within a recently proposed
hierarchical network model using the
susceptible-infected-refractory dynamics of an epidemic. Within
the network model, a population may be characterized by $H$
independent hierarchies or dimensions, each of which consists of
groupings of individuals into layers of subgroups. Detailed
numerical simulations reveal that for $H>1$, global spreading
results regardless of the degree of homophily of the individuals
forming a social circle.  For $H=1$, a transition from global to
local spread occurs as the population becomes decomposed into
increasingly homophilous groups.  Multiple dimensions in
classifying individuals (nodes) thus make a society (computer
network) highly susceptible to large scale outbreaks of infectious
diseases (viruses).


\end{abstract}
\maketitle

\section{Introduction}

Epidemics of all kinds are costly.  The influenza in the early
20th century and the outbreak of SARS in 2003 claimed precious
lives, and the spreading of viruses on the internet has led to
much economic loss when computer systems of governments and
airlines were attacked.  Recent breakdowns of major power networks
in North America and in Europe also brought great loss and
inconvenience to millions of people.  A less damaging example is,
perhaps, that of the spread of a rumor in a community.
Epidemiological processes, together with percolation and problems
on searchability, are the most interesting dynamical problems in a
system of connected entities
\cite{diekmann,albert,dorogovtsev,dorogovtsev1}. There are two key
factors in the study of epidemic spreading, namely how infection
and recovery occur {\em and} how people or computers are
connected. The latter is best described in terms of networks or
graphs \cite{albert,dorogovtsev,dorogovtsev1} in which the nodes
representing people, communities, power plants, or computers are
connected by links representing the acquaintance among people or
the connections between computers. The most important question is
the extent to which a disease (virus) may propagate. Since
infections take place through direct contact between infected (I)
and susceptible (S) nodes, the underlying structure of the network
understandably plays a determining role on the spread of diseases
\cite{Satorras}.

The Susceptible-Infected-Refractory (or recovered/removal) (SIR)
model \cite{diekmann,anderson,murray} has been widely used in
different forms for studying epidemiological processes such as the
spread of influenza (see, for example, Ref. \cite{liu}) and SARS
\cite{riley,lipsitch}, and also for the spread of rumors
\cite{zanette,zanette1}. In its simplest form, a node takes on one
of three states, namely S, I or R. The R-state represents the case
in which an infected node becomes recovered and immunized or,
unfortunately dead. For computer networks, the R-state may
represent recovery with the virus removed and anti-virus software
installed. A node in the R-state is not infectious and will not
become infected again. The underlying connectivity obviously plays
a decisive role. For nodes connected randomly, a substantial
proportion becomes infected and eventually recovered from an
epidemic \cite{sudbury}. However, when a regular network is
gradually re-wired into a small-world network
\cite{albert,dorogovtsev,dorogovtsev1,watts1,watts2} and then a
random network, it was found that a threshold exists on the degree
of disorderness below which the disease can only spread locally,
but not globally \cite{zanette,zanette1}.

Networks have fascinating structural and dynamical properties, and
they have wide applications in physical, biological, social,
economic, and computer systems
\cite{albert,dorogovtsev,dorogovtsev1,strogatz,newman}. Physicists
have made important contributions to the understanding of networks
in recent years. Structurally, the small-world networks
\cite{watts1,watts2} and the growing scale-free networks
\cite{barabasi} have been extensively studied
\cite{albert,dorogovtsev,dorogovtsev1}. Depending on the model,
the degree distribution may show power law or exponential forms.
As the underlying connectivity plays a decisive role in the
studies of dynamical processes, a good network model that reflects
the structural properties is therefore required for modelling
epidemics or other dynamical problems in a connected population,
or in systems in which the entities are grouped in a hierarchical
fashion as in a population. The hierarchical model of searchable
social networks proposed recently by Watts {\em et al.}
\cite{watts3} captures the essential ingredients of a network
model of connected population and has enjoyed much attention.  The
model is physically transparent in that it is motivated by the
observation that individuals within a population are grouped
according to their function into different dimensions, for
example, their occupation, hobby, home district, etc.  The model
exhibits the phenomena of ``six degrees of separation", which was
discovered by Travers and Milgram in the 1960's \cite{travers} and
has been tested in a world-wide experiment on the internet
\cite{watts4}.

Recent works on networks on robustness of networks under attacks
\cite{shargel} and ways to prevent attacks \cite{cohen} indicate
that, again, the connectivity of a network is important. It is,
therefore, important to study the dynamical aspects of the
successful social network model of Watts {\em et al.}
\cite{watts3}. In the present work, we present detailed results of
numerical simulations on the extent of a spread of disease or
virus in the hierarchical social network model of Watts {\em et
al.} within the SIR epidemic dynamics. It is found that the spread
may be local or global if the population can be partitioned in a
unique way. However, when the partition of the population becomes
multi-dimensional, the result is different from that of the
searchability problem. It is found that the spread is always
global.

The plan of the paper is as follows. In Sec.II, we motivate the
hierarchical structure and describe how links are established
between nodes in a hierarchical structure with a tunable degree of
homophily. In Sec.III, we describe the implementation of the SIR
dynamics and discuss the results on the spread of an epidemic in
populations that can be classified into one and two hierarchies.
Section IV summarizes our findings.

\section{Network model of connected population}

The model of Watts {\em et al.} was motivated by the general
structure in the groupings of individuals in a society as shown in
Fig.1 \cite{watts3}. The highest level can be regarded as a
population of $N$ individuals or nodes.  These $N$ nodes may then
be classified or partitioned into $b$ groups. Each group can
further be divided into $b$ subgroups and so on. After $(l-1)$
divisions the structure has a total of $l$ levels and ends at a
level where an individual belongs to a close functional group of
size $g$, where $g$ typically is of the order of $10^{1}$ to
$10^{2}$. Individuals belonging to the same lowest-level subgroup
have the highest chance of becoming most similar or getting
acquaintance. For example, all physicists in academia can be
classified roughly by their research area, say condensed matter
physics or nuclear physics, and further classified by their
subfields, e.g., magnetism or superconductivity, and then further
grouped by their specific research interests, e.g., topics grouped
into the same session in a conference or with the same code in the
PACS classification scheme. Obviously the divisions are usually
not unique. Physicists may be grouped geographically based on the
region their institution is located, and then the country, state,
county, department, and research groups. A population of
individuals or a collection of nodes may thus be characterized by
$H$ hierarchies or dimensions, each of which takes on the
structure shown in Fig.1. The structure of the network of $N$
nodes is then characterized by the parameters $H$, $b$, $l$, and
$\langle g \rangle$ with $N = \langle g \rangle b^{l-1}$, where
$\langle g \rangle$ is the average size of the lowest-level
functional subgroups.

An important quantity within a hierarchy is the social distance
$x_{ij}$ that measures the similarity between two nodes $i$ and
$j$. For nodes belonging to the same lowest-level subgroup in a
given hierarchy, $x_{ij} =1$, otherwise $x_{ij}$ is the number of
levels from the lowest for which the two nodes belong to the same
group. Hence the largest separation is $l$ in a given hierarchy.
For $H>1$, an important geometrical feature is that for nodes $i$
and $j$ with $x_{ij} =1$ in one hierarchy and $j$ and $k$ with
$x_{jk}=1$ in another, $x_{ik}$ is in general {\em not} small. For
example, it is unlikely that the colleague in the office next door
knows your collaborator in another country. Within the context of
epidemics in a community, the hierarchies may be regarded as
family, relatives, friends, and so on; or neighbors in the same
apartment building, same neighborhood, same district and so on.
While the model was motivated by the structure in groupings within
a society, it is obvious that many well-structured systems, e.g.
computer clusters, are connected in a similar fashion.

Computationally, a social network with a tunable degree of
homophily can be constructed as follows \cite{watts2}. For a
population of $N$ individuals with social structure as shown in
Fig.1, the individuals are first assigned randomly into the lowest
subgroups with an average size $\langle g \rangle$ in each of the
$H$ hierarchies. Links between individuals are established as
follows.  For $H > 1$, a hierarchy is randomly chosen.  Two nodes
$i$ and $j$ are then selected randomly within the chosen
hierarchy. A link, specifying that $i$ and $j$ are friends and
hence become acquaintance, is established with a probability
$P(x_{ij}) = \exp{(-\alpha x_{ij})}/\sum_{n=1}^{l} \exp{(-\alpha
n)}$ depending on $x_{ij}$ of the chosen nodes in the chosen
hierarchy. No duplicated links between $i$ and $j$ are allowed.
Here, the parameter $\alpha$ is a measure of homophily of the
system. The process is then repeated until a mean number of
$\langle z \rangle = \langle g \rangle -1$ links are established
for each node (individual) in the system. This guarantees that for
$\alpha \gg - \ln b$ and $H=1$, only links between nodes with
small separation are probable and the individuals are connected
only to those most similar in character, leading to isolated
subgroups of nodes. For $\alpha = -\ln b$, links between
individuals with any social distance are equally probable and a
random network results, with the notion of similarity between
nodes loses its meaning. For intermediate values of $\alpha$,
i.e., $\alpha \approx 1$, the network shows small-world features.
Here, we aim at studying the extent of an epidemic as a function
of $\alpha$ for systems with $H=1$ and $H=2$, respectively.

\section{Epidemic Modelling and Results}

We model epidemics on the social network model by the standard SIR
dynamics, which is implemented as follows \cite{anderson,murray}.
Initially, all nodes are in the S-state and one node is randomly
chosen for infection. At each timestep, a node $i$ is randomly
chosen among all the infected (I) nodes.  A neighbor (friend) $j$
is then selected randomly among all the neighbors of node $i$,
i.e., those with a link connected to $i$. Note that the effects of
dimensionality $H$ is built in through the construction of links.
If node $j$ is susceptible, it becomes infected and the chosen
node $i$ remains in the I-state; otherwise (i.e., node $j$ is
either I or R) the state of node $j$ remains unchanged and the
chosen node $i$ becomes recovered (R) at the end of the timestep.
As time evolves, the number of R-nodes (S-nodes) increases
(decreases); while the number of I-nodes increases initially and
then eventually drops to zero.  The number of R-nodes at the end
of the epidemic is denoted by $N_{R}$.

We performed extensive numerical simulations to explore the
effects of the structural parameter $\alpha$ and the
dimensionality or number of hierarchies $H$. For each set of
parameters, data are collected over $10^{3}$--$10^{4}$ runs, each
of which corresponds to a different realization of the network and
a different initially infected node.  We take $b=2$, $\langle g
\rangle =10$ and hence $\langle z \rangle =\langle g \rangle
-1=9$.  Results for $H=1$ are shown in Fig.2. The structural
effects on epidemics can be studied via the probability
distribution function $P(N_{R})$ for different values of $\alpha$
and different population sizes $N$. For $\alpha = 10 \gg -\ln 2$,
$P(N_{R})$ is independent of $N$ and peaks at $N_{R} \approx 8$,
with no network having $N_{R} > 25$ (see Fig.2(a)). The
independence on the population size $N$ implies that the spread is
{\em local}, for cases of large $\alpha$.  This is the case of
``regular networks" consisting of isolated subgroups in which only
local spread at the lowest level is possible. Each subgroup can be
regarded as a {\em small} fully connected network. Previous
studies showed that about 79\% of nodes in a large fully-connected
network eventually recovered \cite{sudbury}. Using this result as
an estimation gives $0.79 \langle g \rangle \approx 8$ for the
size of a spread, in good agreement with the peak value observed.
As only the mean $\langle g \rangle$ is fixed, each cluster has a
different $g$ leading to a distribution $P(N_{R})$ that extends to
$N_{R}=25$. Simulations using different values of $\langle g
\rangle$ reveal a corresponding shift in $N_{R}$ at which
$P(N_{R})$ shows a maximum.  For the small-world regime
corresponding to $\alpha \approx 1$ (see Fig.2(b)), $P(N_{R})$ is
again $N$-independent and decays exponentially with runs in which
the spread can be up to $N_{R} \approx 10^{3}$. Geometrically,
there exist some links of longer social distance as $\alpha$
decreases, leading to larger but still localized spreads. As
$\alpha$ decreases towards the random network limit of $\alpha =
-ln 2$, $P(N_{R})$ takes on the form of two disjoint parts (see
Fig.2(c)): (i) one that decays exponentially near $N_{R} \approx
0$ (inset) characterizing the small fraction of runs with
small-scale epidemics due to the existence of isolated clusters
and (ii) a $N$-{\em dependent} gaussian distribution
characterizing {\em global} spreading centered at a value
$\overline{N_{R}}$ that scales linearly with $N$.  Writing
$\overline{N_{R}} = a(\langle g \rangle) N$, we found $a \approx
0.66$ for $\langle g \rangle =10$. Extensive simulations reveal
that $a(\langle g \rangle)$ approaches $0.79$, the value for
fully-connected networks, as $\langle g \rangle$ increases. For
$H=1$, the consequence of an epidemic depends sensitively on the
structural parameter $\alpha$.

Social networks usually correspond to $H>1$. For $H>1$, the
characteristic features of an epidemic is very different from that
in $H=1$. The most striking result is that for $H\neq 1$, an
epidemic can {\em always spread globally} regardless of the degree
of homophily (i.e., value of $\alpha$) in the network. Figure 3
shows the results for $H=2$. For fixed $N$ and $\langle z
\rangle$, $P(N_{R})$ becomes {\em independent} of $\alpha$ and
takes on a form identical to that in Fig.2(c).  The majority of
the weight is represented by the gaussian distribution peaked at
$\overline{N_{R}} \approx a(\langle g \rangle) N$, with $a(\langle
g \rangle =10) \approx 0.66$ and increases for larger $\langle g
\rangle$. This result should be contrasted with that of
searchability on the same social network \cite{watts3}, which
occurs for $H>1$ {\em and} $\alpha > 0$ with increasingly
restricted constraints on $H$ and $\alpha$ as $N$ increases. For
epidemics, $H>1$ alone is sufficient for a global spread of the
disease and the result remains valid for different values of $N$.
The independence on $\alpha$ can be understood by the way that the
links are constructed. Although a larger $H$ implies fewer links
per hierarchy, the chance of nodes $i$ and $j$ being similar with
a shorter $x_{ij}$ in one of the hierarchies is higher and so is
the chance that a link exists between nodes $i$ and $j$, even for
large values of $\alpha$. In other words, nodes $i$ and $j$
unknown to each other in one hierarchy has a chance of becoming
connected when additional dimension(s) is (are) added. Thus an
increase in $H$ has the effect of promoting the spread of a
disease in a population by permeating through different connected
nodes in different hierarchies.  This is consistent with daily
experience that a flu may spread first from a parent to a kid
(linked nodes in a family) and then between the kids in school
(linked nodes in school) and so on. We have also studied the case
of $H=3$ and obtained similar results.

To further illustrate the qualitative difference between $H=1$ and
$H=2$ networks, we show the mean fraction of R-nodes, $r \equiv
N^{-1} \sum_{N_{R}=0}^{N} N_{R} P(N_{R})$, as a function of
$\alpha$ in Fig.4. For $H=1$, the large $\alpha$ ($\alpha \gg 1$)
regime gives $r \sim \langle g \rangle/N \approx 0$ corresponding
to the case in Fig.2(a), while the small $\alpha$ ($\alpha < \ln
b$) regime gives an $N$-independent value of $r \approx 0.66$
corresponding to the case in Fig.2(c).  The fraction $r$ drops
sharply in a small intermediate range of $\alpha$ ($\ln 2 < \alpha
< 2$) corresponding to the small-world regime \cite{watts3}. This
feature is similar to that observed in the spreading of rumors in
a small-world construction \cite{zanette,zanette1}. In this
regime, links between nodes with large dissimilarity exist and
hence $N_{R}/N$ increases as $\alpha$ decreases.  One would expect
that the transition becomes sharper as $N$ increases, as $N_{R}$
is restricted only by the connections and is insensitive to $N$ in
the intermediate regime.  For $H=2$, the behavior is qualitatively
different in that $r$ stays at a higher level for all values of
$\alpha$ and the transition disappears as a result of the effect
of the extra hierarchy for hooking up individuals in a population.

\section{Summary}

In summary, epidemiological processes are studied on a
hierarchical social network within the SIR model.  For population
characterized by $H>1$ independent hierarchies, global spreading
results regardless of the structural parameter $\alpha$ and hence
the homophily of individuals forming a social circle.  This result
should be contrasted with that of searchability problems in that
$H>1$ alone is sufficient for a global spread of a disease within
the SIR dynamics. For $H=1$, a transition from global to local
spread occurs as the population becomes decomposed into
increasingly homophilous groups when the structural parameter
$\alpha$ increases.  Since social networks usually correspond to
$H>1$, it is therefore very difficult to confine an epidemic,
unless the ``dimensionality" of the infected nodes can be reduced
at least temporarily.  One way to achieve this is to quarantine
the infected nodes, which is known to be an effective way in
handling infectious diseases. It will also be interesting to
extend the present study to other models of epidemics such as the
SIS \cite{Satorras} and SIRS \cite{kuperman} models.

\acknowledgments {This work was partly  supported by the National
Natural Science Foundation of China under Grant No. 70471081 and
No. 70371069, by a DFG grant under Grant No. TR 3000/3-3,  and by
a grant from the Research Grants Council of Hong Kong SAR
Government under Grant No. CUHK4241/01P.}

\newpage \centerline {\bf FIGURE CAPTIONS}

\bigskip
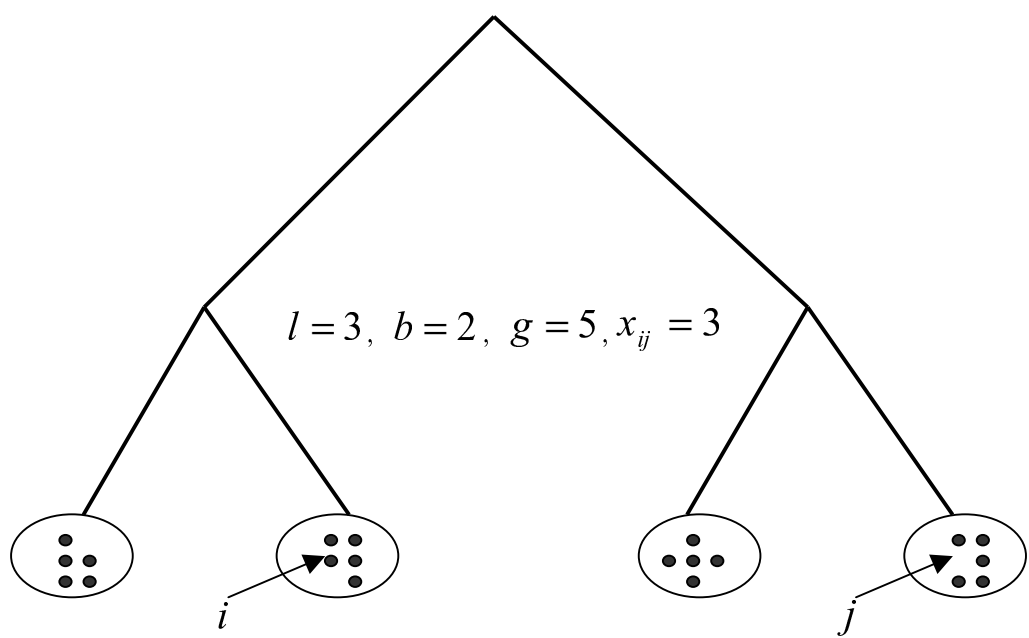
\noindent Figure 1: Schematic diagram of the groupings of individuals
in a hierarchical social network.  A population of $N$ nodes are
classified into $b$ groups.  Each group is further divided into
$b$ subgroups and so on.  After ($l-1$) divisions, the lowest-level
subgroups consist of $g$ individuals.  There $x_{ij}$ is the
social distance between nodes $i$ and $j$.  A population can, in general,
be characterized by $H$ such hierarchies.

\bigskip
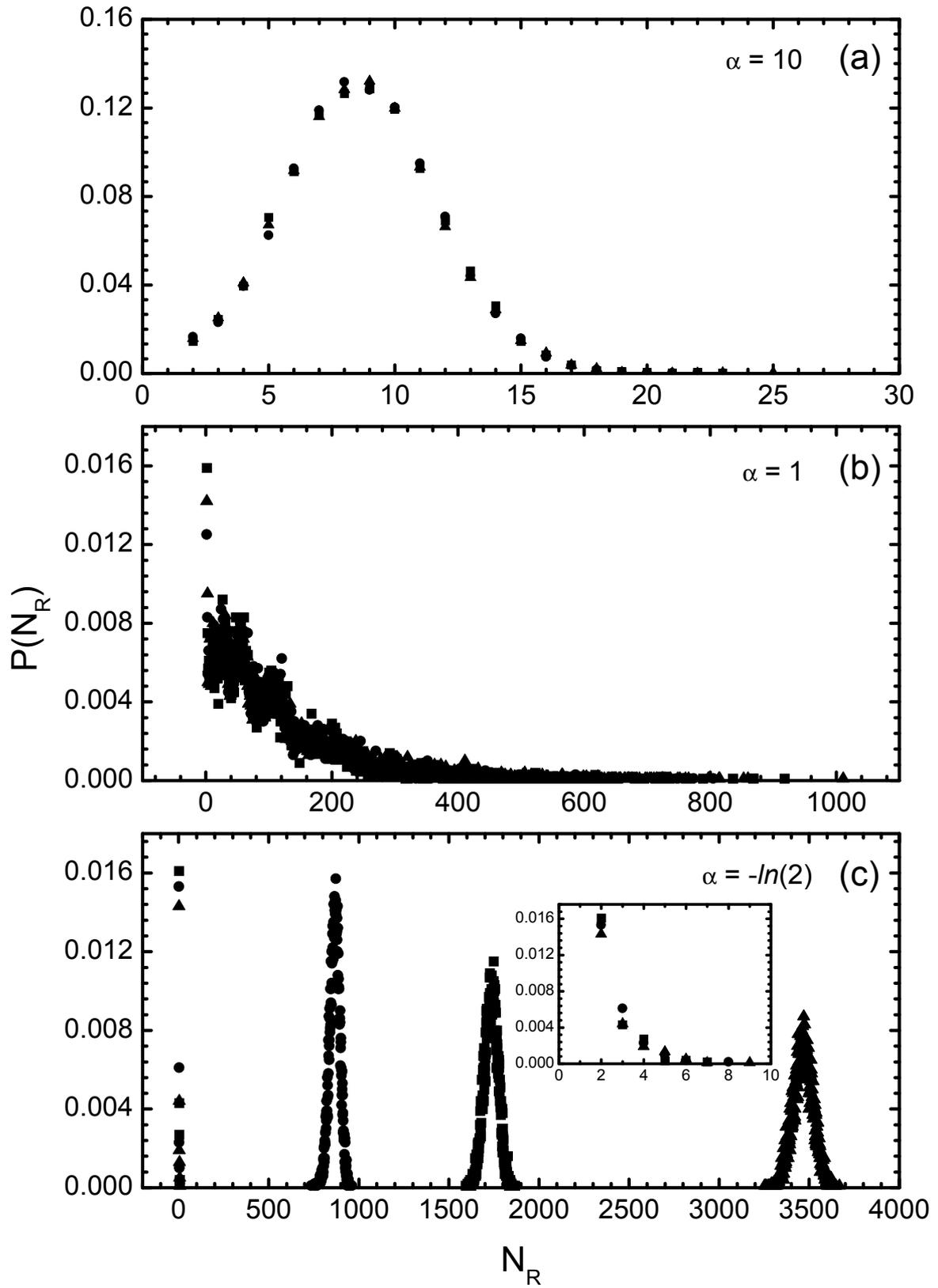
\noindent Figure 2: The probability distribution function $P(N_{R})$
of the number of R-nodes at the end of an epidemic in a population
characterized by $H=1$ hierarchy.   The structural parameter
takes on (a) $\alpha =10$, (b) $\alpha = 1$, (c) $\alpha = - \ln 2$.
Different population sizes $N=1280$ (circles), $2560$ (squares),
and $5120$ (triangles) corresponding to $l=8, 9, 10$ are studied.
Each data point represents an average over $10^{4}$ runs.  The inset
in (c) shows the portion of $P(N_{R})$ for small $N_{R}$.

\bigskip
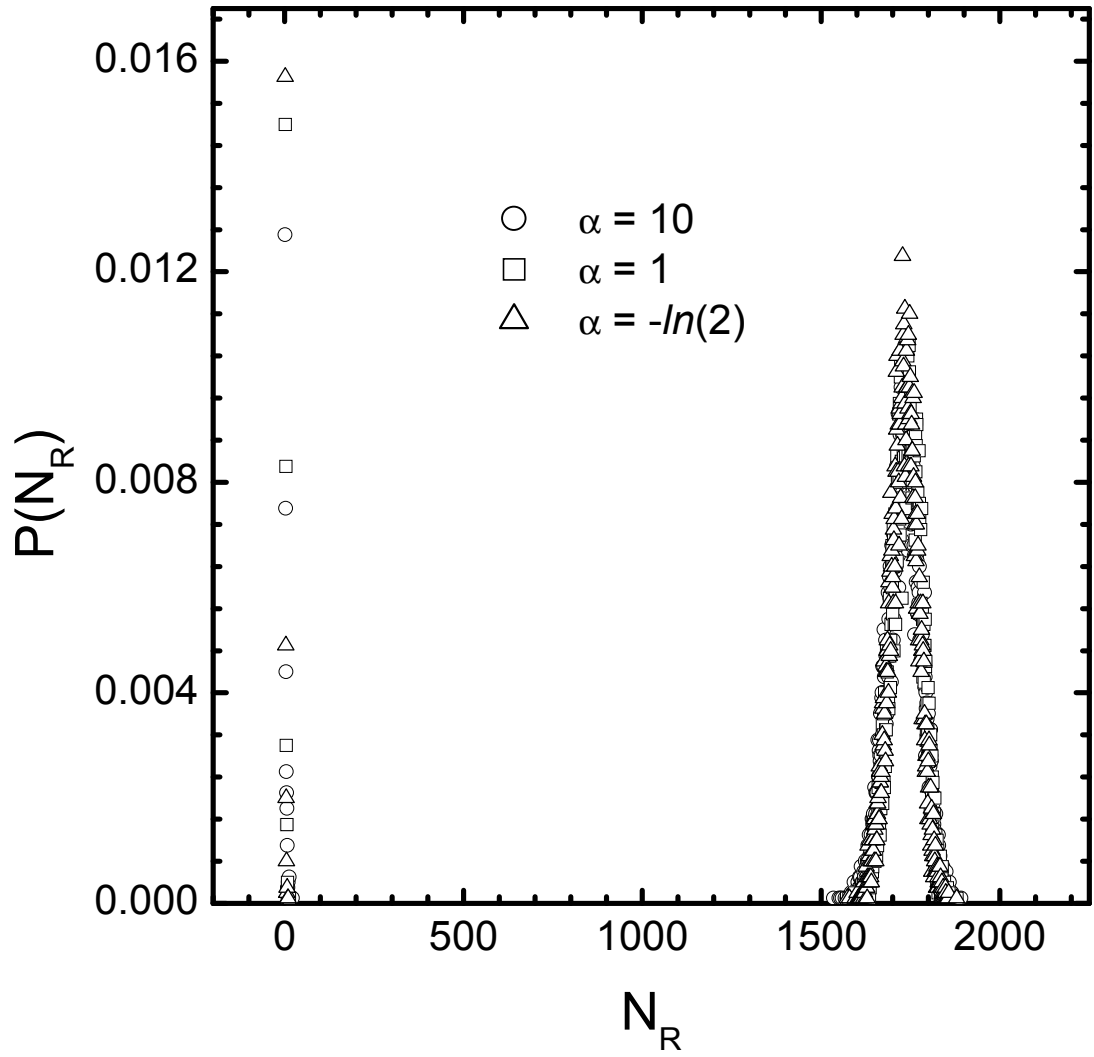
\noindent Figure 3: The probability distribution function $P(N_{R})$
in a population of $N=2560$ ($l=9$) characterized by $H=2$ hierarchies.
The structural parameter takes on $\alpha =10$ (circles),
$\alpha = 1$ (squares), and $\alpha = - \ln 2$ (triangles). Each
data point represents an average over $10^{4}$ runs.  $P(N_{R})$
signifies a global spread of disease regardless of the
structural parameter for $H>1$.

\bigskip
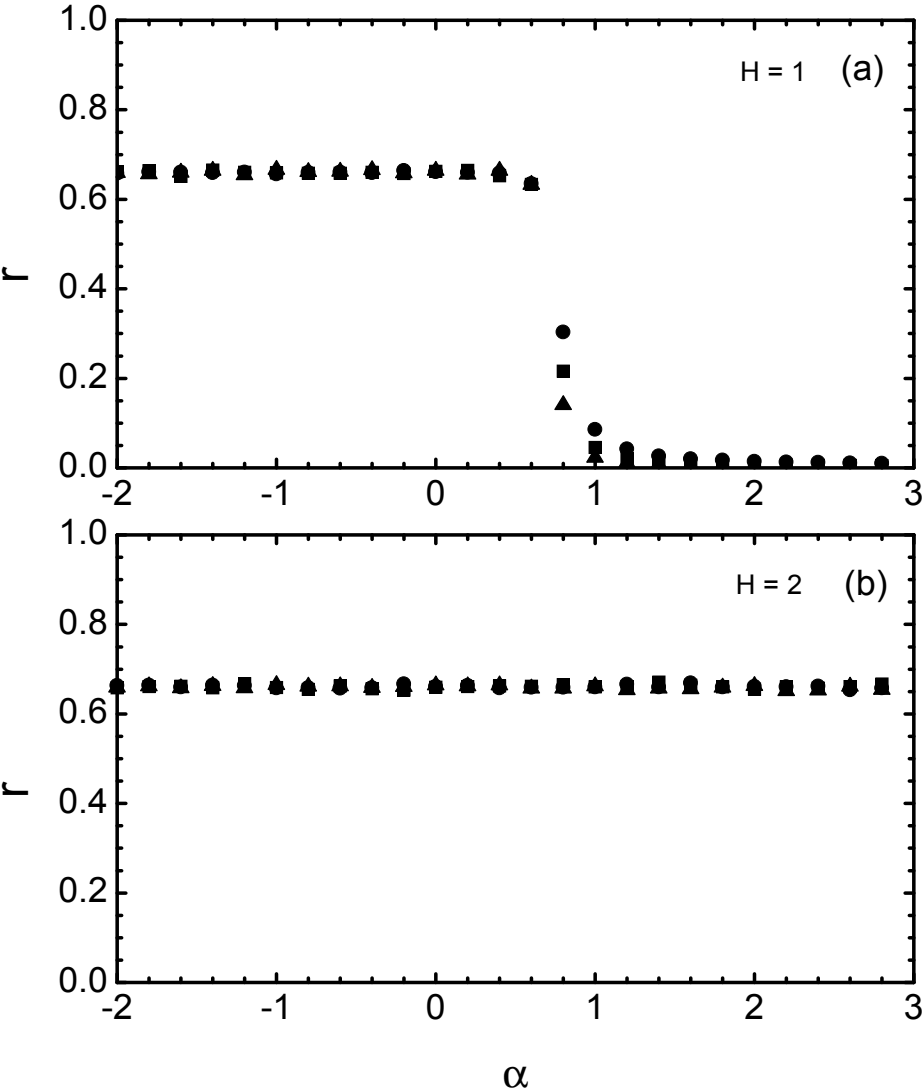
\noindent Figure 4: The fraction of R-nodes at the end of an
epidemic as a function of the structural parameter $\alpha$ in a
population of $N=1280$ (circles), $2560$ (squares), and $5120$
(triangles) characterized by (a) $H=1$ and (b) $H=2$ hierarchies.
Each data point represents an average over $10^{3}$ runs.  For
$H>1$, global spreading occurs for all values of $\alpha$, while a
transition from global to local spreading occurs for $H=1$.

\end{document}